\documentclass[twocolumn,superscriptaddress,aps,prl]{revtex4-1}
\usepackage{amsmath}
\usepackage{graphicx}
\usepackage{dcolumn}
\usepackage{color}
\usepackage{ulem}
\usepackage{bm}
\usepackage{braket}
\usepackage[colorlinks,citecolor=blue]{hyperref}
\usepackage{titlesec}
\begin{document}

\title{Stability of Time-Reversal Symmetry Protected Topological Phases}
\author{Tian-Shu Deng}
\thanks{They contribute equally to this work. }
\affiliation{Institute for Advanced Study, Tsinghua University, Beijing,100084, China}
\author{Lei Pan}
\thanks{They contribute equally to this work. }
\affiliation{Institute for Advanced Study, Tsinghua University, Beijing,100084, China}
\author{Yu Chen}
\email{ychen@gscaep.ac.cn}
\affiliation{Graduate School of China Academy of Engineering Physics, Beijing, 100193, China}
\author{Hui Zhai}
\email{hzhai@tsinghua.edu.cn}
\affiliation{Institute for Advanced Study, Tsinghua University, Beijing,100084, China}
\date{\today}

\begin{abstract}
In a closed system, it is well known that the time-reversal symmetry can lead to Kramers degeneracy and protect nontrivial topological states such as quantum spin Hall insulator. In this letter we address the issue whether these effects are stable against coupling to environment, provided that both environment and the coupling to environment also respect the time-reversal symmetry. By employing a non-Hermitian Hamiltonian with the Langevin noise term and ultilizing the non-Hermitian linear response theory, we show that the spectral functions for Kramers degenerate states can be split by dissipation, and the backscattering between counter-propagating edge states can be induced by dissipation. The latter leads to the absence of accurate quantization of conductance in the case of quantum spin Hall effect. As an example, we demonstrate this concretely with the Kane-Mele model. Our study could also be extended to interacting topological phases protected by the time-reversal symmetry.  

\end{abstract}

\maketitle

Time-reversal symmetry (TRS) protected topological phases, such as TRS protected topological insulator (TI) in two- and three-dimension, are intriguing state-of-matters that have been extensively studied in the past two decades. Unlike the quantum Hall effect, the topological classifications of these states require the presence of TRS \cite{classification_review}. With TRS, the nontrivial topological properties have been firmly established in a closed system \cite{review_TI1,review_TI2}. A natural question is whether the concept of TRS protected topology, as well as its physical consequences such as quantized conductance, still holds in the presence of coupling to environment. This is certainly a very important issue, because for any practical applications of these topological materials, it is inevitable that the materials should be coupled to external environment.

A natural expectation is that the protection from TRS is guaranteed if the Hamiltonian of the system, the environment, and the coupling between the environment and the system all obey TRS. However, this expectation is recently challenged by McGinley and Cooper \cite{Cooper}. They show explicitly that the coupling to environment can lead to de-cohernence between two Kramers doublet state even though the coupling and the environment both obey TRS.  Their argument is deeply rooted in the fact that even if an isolated system obeys TRS, its subsystem can behavior as seemingly violating the TRS. Actually, this fact plays a key role in thermalization of a closed quantum system. In quantum thermalization, considering a pure state of a closed system whose evolution equations obey the TRS, the evolution of its subsystem can undergo an irreversible process that loses information and reach thermalization with the rest of the system acting as a bath \cite{thermalization1,thermalization2}. 

Without loss of generality, we consider an open quantum system coupled to environment through a pair of operators $\hat{\mathcal{O}}$ and $\hat{\mathcal{O}}^\dag$, and both operators obey TRS but do not have to be hermitian. Nevertheless, the entire Hamiltonian, including the system, the system-environment coupling and the environment itself, is hermitian and obeys TRS. By treating the degree-of-freedoms of the environment with the Markovian approximation, the open quantum system can be described by a non-Hermitian Hamiltonian with a Langevin noise term, which ensures the quantum mechanical commutative relation and preserves the trace of the density matrix \cite{supple}. This non-Hermitian Hamiltonian can be generally written as
\begin{equation}
\hat{H}=\hat{H}_0-i\gamma\hat{\mathcal{O}}^\dag\hat{\mathcal{O}}+\hat{\mathcal{O}}^\dag\hat{\xi}+\hat{\xi}^\dag\hat{\mathcal{O}},
\end{equation}  
where $\gamma$ is the dissipation strength. $\hat{H}_0$ is the Hermitian Hamiltonian of the system itself, and it also obeys TRS. $\hat{\xi}$ is the Langevin noise operators that satisfy $\langle\hat{\xi}(t)\hat{\xi}^\dag(t')\rangle=2\gamma\delta(t-t')$ and $\langle\hat{\xi}^\dag(t)\hat{\xi}(t')\rangle=\langle\hat{\xi}(t)\hat{\xi}(t')\rangle=\langle\hat{\xi}^\dag(t)\hat{\xi}^\dag(t')\rangle=0$. All the calculation done with this non-Hermitian calculation should be accompanied by averaging over the Langevin noise term in the end. In Ref. \cite{Pan}, we have developed a non-Hermitian linear response theory. This theory starts with the equilibrium state of $\hat{H}_0$ and treats dissipation order by order, which determines how an equilibrium system responds to a weak dissipation. To implement the non-Hermitian linear response theory, we should introduce an interaction picture which separates out the dissipation term from the system term. For instance, in the interaction picture, we should define $\hat{\mathcal{O}}^I(t)=e^{i\hat{H}_0t}\hat{\mathcal{O}} e^{-i\hat{H}_0 t}$, and similar definitions for other operators with upper scribe $\text{I}$.

\textbf{Summary of Results.} We consider generally a pair of Kramers degenerate eigen-states of $\hat{H}_0$, say $\ket{\Psi}_1$ and $\ket{\Psi}_2$. When we specifically consider a TRS protected TI, they can be chosen as a pair of degenerate edge states located at the same edge. We denote the Hilbert space formed by these two states as $\mathcal{H}_\text{K}$.  In this letter, by studying the linear response of the density matrix, the Green's function and the matrix element of a local impurity potential respectively, we obtain three main results as summarized below:

\textit{1. Loss of Coherence.} Suppose that initially the quantum state is a pure state in $\mathcal{H}_\text{K}$ as $\ket{\Psi}=\alpha_1\ket{\Psi_1}+\alpha_2\ket{\Psi_2}$, where $\alpha_{i=1,2}$ are two constants, the initial density matrix is given by $\hat{\rho}_\text{K}(0)=|\Psi\rangle\langle\Psi|$. By turning on the dissipation, the quantum state evolves under $\hat{H}$ and the density matrix becomes $\hat{\rho}(t)$. By projecting onto $\mathcal{H}_\text{K}$ space by the projection operator $\hat{\Pi}_\text{K}$, one can obtain the projected density matrix $\hat{\rho}_\text{K}(t)=\frac{1}{\mathcal{N}}\hat{\Pi}_\text{K}\hat{\rho}(t)\hat{\Pi}_\text{K}$, with normalization factor $\mathcal{N}={\rm Tr}(\hat{\Pi}_\text{K}\hat{\rho}(t)\hat{\Pi}_\text{K})$. We define $\delta \hat{\rho}_\text{K}(t)=\hat{\rho}_\text{K}(t)-\hat{\rho}_\text{K}(0)$. We show that $\delta \hat{\rho}_\text{K}(t)$ is not proportional to $\hat{\rho}_\text{K}(0)$.

\textit{2. Break of Degeneracy.} With dissipation, the retarded Green's function in $\mathcal{H}_\text{K}$ space is a two-by-two matrix $\mathcal{G}$ with the matrix elements defined as  
\begin{equation}
\mathcal{G}_{ij}=-i\Theta(t)\langle\{\hat{c}_i(t),\hat{c}^{\dag}_{j}(0)\}\rangle, \label{greenf}
\end{equation}
where $\hat{c}_i$ and $\hat{c}^\dag_i$ are annihilation and creation operators corresponding to eigenstates $\ket{\Psi_i}$ of $\hat{H}_0$, and $\hat{c}_i(t)=e^{i\hat{H} t}\hat{c}_ie^{-i\hat{H}t}$. We show that $\mathcal{G}_{ij}$ is no longer proportional to an identity matrix in the $\mathcal{H}_\text{K}$ space. 

\textit{3. Presence of Backscattering.} For a local impurity potential $\hat{V}$, we consider the matrix element of this impurity potential between two Kramers states, i.e. $V^0_{ij}=\bra{\Psi_i} \hat{V}\ket{\Psi_j}$. Suppose, without dissipation, this matrix element is identically zero for $i\neq j$. This can be satisfied, for instance, when $\ket{\Psi_1}$ and $\ket{\Psi_2}$ are respectively the left-moving and the right-moving edge states of a quantum spin Hall state. With dissipation, we need to consider
\begin{equation}
V_{ij}(t)=\bra{\Psi_i}\hat{V}(t)\ket{\Psi_j}, \label{Tmatrix}
\end{equation}
where $\hat{V}(t)=e^{i\hat{H} t}\hat{V}e^{-i\hat{H}t}$. We show that $V_{ij}(t)\neq 0$ for $i\neq j$. 

Result \textbf{1} naturally leads to $S_\text{v}(t)\neq S_\text{v}(t=0)$ and $S_\text{R}(t) \neq S_\text{R}(t=0)$, where $S_\text{v}(t)=-\text{Tr}\hat{\rho}_\text{K}(t)\log\hat{\rho}_\text{K}(t)$ is the von Neumann entropy and $S_\text{R}(t)=-\log\text{Tr}\hat{\rho}^2_\text{K}(t)$ is the second Renyi entropy. That is to says, for $t>0$, the entropy becomes non-zero and the system loses its phase coherence. This is consistent with the result presented in Ref. \cite{Cooper}. Result \textbf{2} and \textbf{3} are the central results of this work. With the Result \textbf{2}, we can further plot the spectrum function $A(\omega)$, which shows two split peaks. This means the lack of Kramers degeneracy for a non-Hermitian open system even though the coupling to environment also respects the TRS. Result \textbf{3} is directly related to the two-dimensional quantum spin Hall. Quantized conductance is the hallmark of quantum spin Hall, due to the forbidden of the backscattering between the left- and the right-moving edge states. Therefore, the presence of backscattering means that the conductance of a quantum spin Hall state is not perfectly quantized. This physics has been qualitatively discussed in Ref. \cite{dissipation,Cooper}. This is perhaps one of the reasons that accuracy of quantization observed in quantum spin Hall samples so far \cite{exp,dissipation,Du_exp} is far less than that observed in quantum Hall samples, in addition to other possible explanation such as the inelastic scatterings \cite{exp}. These results are essentially due to the irreversible nature of the bath, and bare a lot of similarity as the H-theorem in statistical mechanics. In other word, these results can be viewed as the manifestations of the H-theorem in the TRS protected topology.   

\textbf{Application of the Schur's Lemma.} Before proceeding into the details of the derivation, we should emphasize that these results essentially rely on the TRS being an anti-unitary symmetry. In other words, if the symmetry that protects the topological phase is a unitary symmetry, the phenomena \textbf{1}-\textbf{3} described above should not occur. Mathematically, the difference roots in the celebrated Schur's Lemma in the group theory \cite{Schur}. The Schur's Lemma says that, for a unitary group, if an operator $\hat{M}$ commutes with all elements in the group, then this operator, in an irreducible representation, has to be proportional to an identity matrix. Nevertheless, when the Schur's Lemma is applied to an anti-unitary group, not only the operator $\hat{M}$ has to commute with all elements in the group, but also the operator $\hat{M}$ has to be a hermitian operator, then this operator is proportional to an identity matrix in an irreducible representation \cite{anti-Schur}. As we will see below, when the hermitian condition and respecting the anti-unitary symmetry condition cannot be satisfied simultaneously, this operator is generally no longer proportional to identity. This is the key mathematical reason responsible for the difference between the unitary symmetry protection and the anti-unitary symmetry protection.

To be more concrete, we will give two examples that will be repeated used below:

The first example is about $\hat{\Pi}_\text{K}\hat{\mathcal{O}}^\text{I}(t)\hat{\mathcal{O}}^{\dag,\text{I}}(t)\hat{\Pi}_\text{K}$. By using the fact that the states in $\mathcal{H}_\text{K}$ are degenerate state of $\hat{H}_0$, we have $\hat{\Pi}_\text{K}e^{\pm i\hat{H}_0t}=e^{\pm i\hat{H}_0t}\hat{\Pi}_\text{K}=e^{\pm iE_0t}\hat{\Pi}_\text{K}$, and therefore, $\hat{\Pi}_\text{K}\hat{\mathcal{O}}^\text{I}(t)\hat{\mathcal{O}}^{\dag,\text{I}}(t)\hat{\Pi}_\text{K}=\hat{\Pi}_\text{K}\hat{\mathcal{O}}\hat{\mathcal{O}}^\dag\hat{\Pi}_\text{K}$. Note that $\hat{\Pi}_\text{K}\hat{\mathcal{O}}\hat{\mathcal{O}}^\dag\hat{\Pi}_\text{K}$ is hermitian and time-reversal symmetric. It is also important to note that the Hilbert space $\mathcal{H}_\text{K}$ of two Kramers degenerate states forms an irreducible representation of the TRS, thus, the projection $\hat{\Pi}_\text{K}$ enforces the restriction to an irreducible space of TRS. Therefore, this term obeys the Schur's Lamma, and consequently, it is proportional to identity. The same holds for $\hat{\Pi}_\text{K}\hat{\mathcal{O}}^{\dag,\text{I}}(t)\hat{\mathcal{O}}^{\text{I}}(t)\hat{\Pi}_\text{K}$. 

The second example is about $\hat{\Pi}_\text{K}\hat{\mathcal{O}}^\text{I}(t)\hat{\Pi}_\text{K}$ and $\hat{\Pi}_\text{K}\hat{\mathcal{O}}^{\dag,\text{I}}(t)\hat{\Pi}_\text{K}$. It can also be shown that $\hat{\Pi}_\text{K}\hat{\mathcal{O}}^\text{I}(t)\hat{\Pi}_\text{K}=\hat{\Pi}_\text{K}\hat{\mathcal{O}}\hat{\Pi}_\text{K}$. 
Because $\hat{\mathcal{O}}$ is time-reversal symmetric but is generally not hermitian, this term does not satisfy the Schur's Lemma for the anti-unitary TRS. The same holds for $\hat{\Pi}_\text{K}\hat{\mathcal{O}}^{\dag,\text{I}}(t)\hat{\Pi}_\text{K}$. However, for unitary symmetry, because the Schur's Lemma does not require the operator being hermitian, all these operators are  proportional to identity in an irreducible space if they obey the unitary symmetry. Hence, the unitary symmetry protected topological states are stable against coupling to environment. 

\textbf{Loss of Coherence.} Here we consider density matrix in the interaction picture $\hat{\rho}(t)=\hat{\cal U}(t)\hat{\rho}_\text{K}(0)\hat{\cal U}^\dag(t)$ with $\hat{\cal U}(t)=e^{i\hat{H}_0t}e^{-i\hat{H}t}$, and by expanding $\hat{\rho}(t)$ to the leading order of $\gamma$ and averaging the noise, we can obtain
\begin{align}
& \hat{\rho}(t)-\hat{\rho}_\text{K}(0)= \nonumber\\
&2\gamma \int_0^t\left(-\frac{1}{2}\{\hat{\mathcal{O}}^{\dag,\text{I}}(t^\prime)\hat{\mathcal{O}}^\text{I}(t^\prime),\hat{\rho}_{\text{K}}(0)\}+\hat{\mathcal{O}}^\text{I}(t^\prime)\hat{\rho}_{\text{K}}(0)\hat{\mathcal{O}}^{\dag,\text{I}}(t^\prime)\right)dt^\prime, \label{deltarho}
\end{align}
where the second term results from averaging over the Langevin noise. By projecting back to $\mathcal{H}_\text{K}$, the first term in the r.h.s of Eq. \eqref{deltarho} can be written as 
\begin{align}
\{\hat{\Pi}_\text{K}\hat{\mathcal{O}}^\text{I}(t^\prime)\hat{\mathcal{O}}^{\dag,\text{I}}(t^\prime)\hat{\Pi}_\text{K},\hat{\rho}_{\text{K}}(0)\},\label{rho1}
\end{align}
and the second term in the r.h.s of Eq. \eqref{deltarho} can be written as
\begin{align}
\left[\hat{\Pi}_\text{K}\hat{\mathcal{O}}^\text{I}(t^\prime)\hat{\Pi}_\text{K}\right]\hat{\rho}_\text{K}(0)\left[\hat{\Pi}_\text{K}\hat{\mathcal{O}}^{\dag,\text{I}}(t^\prime)\hat{\Pi}_\text{K}\right]\label{rho2}
\end{align}
With the two examples discussed above, we can conclude that Eq. \eqref{rho1} is proportional to $\hat{\rho}_\text{K}(0)$ but Eq.\eqref{rho2} is not proportional to $\hat{\rho}_\text{K}(0)$. Hence $\delta\hat{\rho}_\text{K}(t)$ is not proportional to $\hat{\rho}_\text{K}(0)$, and the entropy changes. 

\textbf{Break of Degeneracy.} Here we apply the linear response theory to the Green's function defined in Eq.\eqref{greenf}, and consider that the Kramers doublet are both occupied by a pair of fermions. Similar as the discussion above, we consider $\delta G_{ij}=\mathcal{G}_{ij}-\mathcal{G}^{(0)}_{ij}$, where 
\begin{equation}
\mathcal{G}^{(0)}_{ij}=-i\Theta(t)\langle\{\hat{c}^\text{I}_i(t),\hat{c}^{\dag,\text{I}}_{j}(0)\}\rangle,
\end{equation}
and $\mathcal{G}^{(0)}_{ij}$ is the Green's function without dissipation. 
It is easy to see that $\mathcal{G}^0_{ij}\propto\delta_{ij}$. What we need to show is that $\delta\mathcal{G}_{ij}$ is not proportional to $\delta_{ij}$. We can also expand $\delta\mathcal{G}_{ij}$ to the leading order of $\gamma$. We shall not show the full expression of this term here \cite{supple}. Generally speaking, there are two types of terms in the leading order expansion. One kind of terms include, for instance, 
\begin{equation}
\int_{0}^t \left\langle\hat{c}^{\dag,\text{I}}_j(0)\left[\hat{\Pi}_\text{K}\hat{\mathcal{O}}^{\dag,\text{I}}(t_1)\hat{\mathcal{O}}^\text{I}(t_1)\hat{\Pi}_\text{K}\right]\hat{c}^\text{I}_i(t)\right\rangle dt_1,
\end{equation}
which involve $\hat{\Pi}_\text{K}\hat{\mathcal{O}}^{\dag,\text{I}}(t_1)\hat{\mathcal{O}}^{\text{I}}(t_1)\hat{\Pi}_\text{K}$. The other kind of terms include, for instance,   
\begin{equation}
\int_{0}^t \left\langle\hat{c}^{\dag,\text{I}}_j(0)\left[\hat{\Pi}_\text{K}\hat{\mathcal{O}}^{\dag,\text{I}}(t_1)\hat{\Pi}_\text{K}\right]\hat{c}^\text{I}_i(t)\left[\hat{\Pi}_\text{K}\hat{\mathcal{O}}^\text{I}(t_1)\hat{\Pi}_\text{K}\right]\right\rangle dt_1
\end{equation}
which involve $\hat{\Pi}_\text{K}\hat{\mathcal{O}}^\text{I}(t_1)\hat{\Pi}_\text{K}$ and $\hat{\Pi}_\text{K}\hat{\mathcal{O}}^{\dag,\text{I}}(t_1)\hat{\Pi}_\text{K}$. 
With the two examples discussed above, we can also see that the first kind of term is still proportional to $\delta_{ij}$ but the second kind of term is not. Hence, up to the leading order of $\gamma$, $\mathcal{G}$ is already not an identity matrix and the spectrum is split. 
 
\textbf{Presence of Backscattering.} Here we consider the matrix element defined in Eq. \eqref{Tmatrix}. Similarly, we define $\delta V_{ij}=V_{ij}(t)-V^0_{ij}$, and we expand $\delta V_{ij}$ to the leading order of $\gamma$ \cite{supple}. Here, as a typical example, we focus on one of the terms that are similar as the ones discussed above, which reads
\begin{align}
&\int_{0}^t \bra{\Psi_i}\left[\hat{\Pi}_\text{K}\hat{\mathcal{O}}^{\dag,\text{I}}(t_1)\hat{V}^\text{I}(t)\hat{\mathcal{O}}^\text{I}(t_1)\hat{\Pi}_\text{K}\right] \ket{\Psi_j} dt_1 \nonumber \\
=&\int_{0}^t \bra{\Psi_i}\left[\hat{\Pi}_\text{K}\hat{\mathcal{O}}^{\dag}e^{i\hat{H}_0(t-t_1)}\hat{V}e^{-i\hat{H}_0(t-t_1)}\hat{\mathcal{O}}\hat{\Pi}_\text{K}\right] \ket{\Psi_j} dt_1 \label{Tleading02}
\end{align}
Here we should note a difference between the discussion here and the two cases above. Above two results can both be proved within the Kramers degenerate space $\mathcal{H}_\text{K}$. However, if in this case we are restricted in the $\mathcal{H}_\text{K}$ space, $V^\text{I}$ is an identity matrix that commutes with $\hat{H}_0$. Thus, Eq. \eqref{Tleading02} becomes 
\begin{equation}
\int_{0}^t\bra{\Psi_i}\left[\hat{\Pi}_\text{K}\hat{\mathcal{O}}^{\dag}\hat{\mathcal{O}}\hat{\Pi}_\text{K}\right] \ket{\Psi_j} dt_1, 
\end{equation}
where $\hat{\Pi}_\text{K}\hat{\mathcal{O}}^{\dag}\hat{\mathcal{O}}\hat{\Pi}_\text{K}$ is hermitian and obeys TRS. One can show that this holds for other terms in the leading order expansion of $\delta V_{ij}$. Therefore, restricted in $\mathcal{H}_\text{K}$ space, it is an identity matrix and cannot induce backscattering. 

Hence, we should consider $\hat{V}$ operator out of $\mathcal{H}_\text{K}$ space, where $\hat{V}$ is no longer represented as identity and in general does not commute with $\hat{H}_0$. Then, it is easy to see that the operator in the $[...]$ of Eq. \eqref{Tleading02} does not respect TRS, although it is a hermitian one. Therefore, this term does not obey the Schur's lemma and is not proportional to identity. Once not being identity matrix, nothing guarantees this term to be a diagonal matrix, and generically, the off-diagonal matrix elements exist, which lead to $\delta V_{ij}\neq 0$ for $i\neq j$. Similar discussions can be applied to other terms in the first order expansion of $\delta V_{ij}$. Taking $i$ and $j$ as a pair of degenerate counter-propagating edge states of a quantum spin Hall, we have now established the presence of backscattering and the absence of perfect quantization of conductance.

\textbf{Example: the Kane-Mele model with Dissipation.} Here we use the celebrated Kane-Mele model on a honeycomb lattice for two-dimensional quantum spin Hall to illustrate these three results more concretely \cite{kane-mele}. For this model, we have
\begin{align}
&\hat{H}_{0}=J\sum_{\langle i,j\rangle,s}\hat{c}_{i,s}^{\dagger}\hat{c}_{j,s}+i\lambda_{\text{SO}}\sum_{\langle\langle i,j\rangle\rangle,s,s^\prime}\nu_{ij}\hat{c}_{i,s}^{\dag}\sigma^{z}_{ss^\prime}\hat{c}_{j,s\prime}\nonumber\\
&+i\lambda_{\text R}\sum_{\langle i,j\rangle,s,s^\prime}\hat{c}_{i,s}^{\dagger}(\bm{\sigma}\times\bm{d}_{ij})^{z}_{ss^\prime}\hat{c}_{j,s^\prime}+\lambda_{\nu}\sum_{i,s}\xi_{i}\hat{c}_{i,s}^{\dagger}\hat{c}_{i,s}.
\end{align}
where $i$ and $j$ are site index and $s$ and $s^\prime$ are spin index, and $\bm{\sigma}$ are the Pauli matrices. 
The first term is the nearest neighbor hopping with strength $J$. The second term is a spin-orbit coupling between second neighbor hopping, with $\nu_{ij}=\pm 1$ and strength $\lambda_\text{SO}$. The third term is the nearest Rashba term with strength $\lambda_\text{R}$, where $\bm{d}_{ij}$ is the vector connecting $i$- and $j$-sites. The last term is a staggered potential with $\xi_i=\pm 1$ for different sublattices and strength $\lambda_{\nu}$. We choose the parameters such that the model is in the topological nontrivial insulator state. 

\begin{figure}[t] 
\centering
\includegraphics[width=0.42\textwidth]{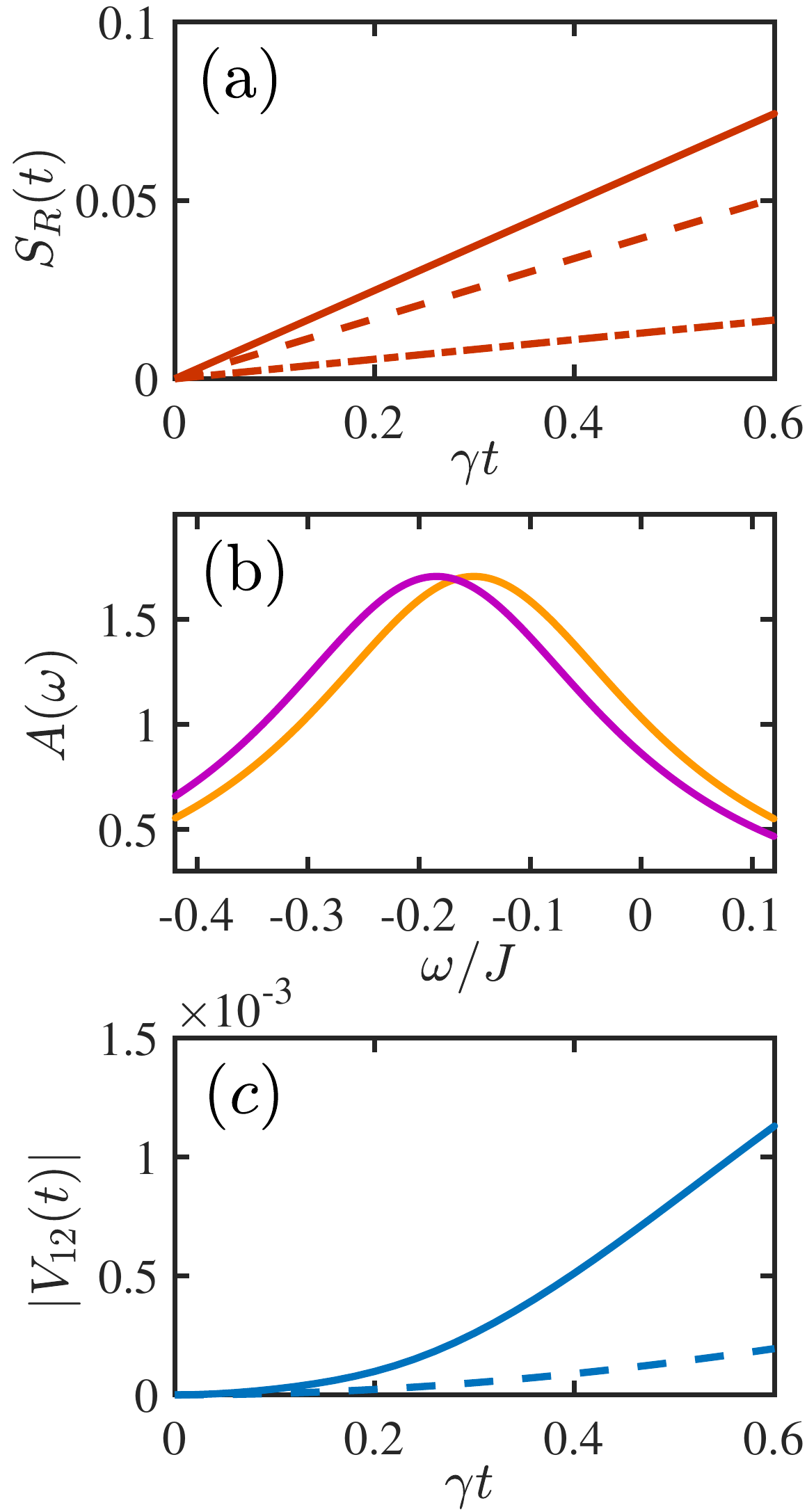}
\caption{(a) The linear response of second Renyi entropy $S_\text{R}(t)$ as a function of time. Here $M=30$ (solid line), $=20$ (dashed line) and $=10$ (dotted-dashed line), respectively. (b) The spectral function $A(\omega)$ for two Kramers degenerate states with dissipation, with $M=20$. Without dissipation, the eigen-energies of these two states are degenerate and equal $-0.17J$. 
(c) Time evolution of the matrix element of an impurity potential $V_{12}(t)$ between two degenerate edge states with $M=20$. The solid line includes contributions from all states and the dashed line only includes contributions from the edge states. In all plots, the size of honeycomb lattice is set as $N_x=8$ and $N_y=30$. The dissipation strength $\gamma$ is taken as $0.2J$, and other parameters in the Kane-Mele model are chosen as $\lambda_{\text{SO}}/J=0.06$, $\lambda_{R}/J=0.05$ and $\lambda_\nu/J=0.1$. The impurity strength $V$ is taken as $V=J$. }
     \label{kane_mele}
\end{figure}

We introduce the coupling operator $\hat{O}$ either defined on site-$i$ as 
\begin{align}
\hat{\mathcal{O}}=i\sum_{s,s'}\hat{c}^\dagger_{i,s}\sigma_{ss^\prime}^y\hat{c}_{i,s^\prime}, 
\hat{\mathcal{O}}^\dagger=-i\sum_{s,s^\prime}\hat{c}^\dagger_{i,s}\sigma^y_{ss^\prime}\hat{c}_{i,s^\prime},
\end{align}
or defined on a nearest neighboring link $\langle i,j\rangle$ as
\begin{align}
\hat{\mathcal{O}}=\sum_{s}\hat{c}^\dagger_{i,s}\hat{c}_{j,s}, 
\hat{\mathcal{O}}^\dagger=\sum_{s}\hat{c}^\dagger_{j,s}\hat{c}_{i,s}.
\end{align}
It is easy to see that the operators defined above obey TRS and are not hermitian. In the numerical simulation, we include a number of $\hat{O}$ operators defined above located at the edge of an sample, and the number of coupling operators is denoted by $M$. We note that the discussion above can be generalized straightforwardly to the cases with more coupling operators. 

Here we numerically diagnolize the Kane-Mele model on a $N_x\times N_y$ sample, with open boundary condition along $\hat{x}$ and periodical boundary condition along $\hat{y}$. Here we should emphasize that, in order to obey TRS, the operators $\hat{\mathcal{O}}$ have to be a quadratic fermion operator, and therefore, the total Hamiltonian contains four-fermion terms and cannot be solved by diagnolizing a quadratic matrix. Therefore, although the spectrum of $\hat{H}_0$ can be obtained exactly, the effects of dissipation still needs to be computed by the non-Hermitian linear response theory, and the numerical results are shown in Fig. \ref{kane_mele}. We take two edge states of $\hat{H}_0$ with same energy and located at the same edge as the Kramers degenerate states $\ket{\Psi_i}$ ($i=1,2$). First, starting with an initial pure state $\ket{\Psi}=\frac{1}{\sqrt{2}}\ket{\Psi_1}+\frac{1}{\sqrt{2}}\ket{\Psi_2}$, we determine the evolution of density matrix, with which we compute the time dependence of the second Renyi entropy $S_\text{R}(t)$ shown in Fig. \ref{kane_mele}(a). One can see that the entropy increases linearly in time, with larger slop for larger $M$. Secondly, the spectral function $A(\omega)$ for these two states are shown in Fig. \ref{kane_mele}(b). One can see that the peaks of two spectral functions are split. Thirdly, we compute the backscattering matrix element of an on-site impurity potential between these two edge states. As discussed above, this calculation needs to involve the contributions out of the Kramers degenerate states. Here we plot the results with contributions from all states, as well as results with contributions from edge states only, which shows both edge and bulk states contribute to the non-zero matrix elements. 

\textit{Remarks.} In summary, we have discussed how a system responds to dissipations, with TRS imposed on both the system and the environment, as well as the coupling operators between them. The main results are the absence of Kramers degeneracy and the absence of accurate quantization of conductance for TRS protected TI. We also recover the results reported in Ref. \cite{Cooper} on losing of phase coherence. However, different from Ref. \cite{Cooper}, we employ a non-Hermitian Hamiltonian formalism for open quantum system with the Markovian approximation to the environment. Different from many works on non-Hermitian physics in recent literatures, our non-Hermitian Hamiltonian contains Langevin noise term to ensure unitarity. In fact, above discussions show that the Langevin noise terms plays a crucial role because most terms violating the Schur's Lemma are essentially from the Langevin noise average. We should also emphasize that the way we impose the TRS symmetry leads to quartic non-Hermitian terms. This also makes our model different from those considered in recent works on topological classification of non-Hermitian Hamiltonian, where the models under consideration are always quadratic \cite{classification_non_Hermitian,classification_non_Hermitian2}.

\textit{Acknowledgment.} We thank Zhong Wang, Yun Li, Chaoming Jian, Hong Yao and Pengfei Zhang for helpful discussions. This work is supported by Beijing Outstanding Young Scientist Program (HZ), NSFC Grant No. 11734010 (HZ and YC), NSFC under Grant No. 11604225 (YC), MOST under Grant No. 2016YFA0301600 (HZ) and Beijing Natural Science Foundation (Z180013) (YC).

\setcounter{equation}{0}
\setcounter{figure}{0}
\setcounter{table}{0}
\setcounter{section}{0}
\renewcommand{\theequation}{S\arabic{equation}}
\renewcommand{\thefigure}{S\arabic{figure}}
\renewcommand{\thesection}{S\arabic{section}}
\onecolumngrid
\flushbottom
\newpage

\subsection*{\normalsize Supplementary Material for ``Stability of Time-Reversal Symmetry Protected Topological Phases"}

\section{I. The Equivalence of Total Hamiltonian and Non-Hermitian Effective Hamiltonian} \label{sec1}

In this section, we will demonstrate the equivalence between a total hamiltonian including the bath and the effective non-Hermitian hamiltonian with the Langevin noise term for the system only. Specifically, we will prove that any operator evolution with the total hamiltonian after averaging over the environment is equivalent to the operator evolution by effective non-Hermitian hamiltonian after averaging over the Langevin noise. To start with, we consider the total hamiltonian as
\begin{align}
&\hat{H}_{\rm tot}=\hat{H}_{\rm 0}+\hat{H}_{\rm B}+\hat{H}_{\rm int},\\
&\hat{H}_{\rm B}=\sum_{\alpha,m}\omega^{}_{\alpha} \hat{a}^\dagger_{\alpha,m} \hat{a}_{\alpha,m},\\
&\hat{H}_{\rm int}=\sum_{\alpha,m}\left(g^{}_{\alpha,m}\hat{\cal O}_m^\dag \hat{a}^{}_{\alpha,m}+g^{}_{\alpha,m}\hat{a}^\dag_{\alpha,m} \hat{\cal O}_m^{}\right),\label{Ham0}
\end{align}
where $\hat{H}_{\rm 0}$, $\hat{H}_{\rm B}$ are the Hamiltonian of system and bath respectively, and $\hat{H}_{\rm int}$ denotes the interaction between them.  $\hat{a}_{\alpha,m}$ is annihilation operators for harmonic oscillators of mode $\alpha$ and channel $m$. $\hat{a}_{\alpha, m}$ is an operator acting on the environment Hilbert space, satisfying  $\left[\hat{a}_{\alpha,m}, \hat{a}_{\beta,n}^{\dagger}\right]=\delta_{\alpha\beta}\delta_{mn}$. $\hat{\cal O}_m$ is the operator acting on system and $g_{\alpha, m}$ is the coupling strength between $\hat{\cal O}_m$ and $\hat{a}_{\alpha, m}$. Since $\hat{a}=\hat{x}-i\hat{p}$ is invariant under time-reversal symmetry (TRS) transformation, here we impose that $\hat{\cal O}_m$ is TRS. Further we assume the environment is at zero temperature, then we have
\begin{equation}
\langle \hat{a}_{\alpha,m}(t)\hat{a}^{}_{\beta,n}(t_1)\rangle_{\rm B}=\langle \hat{a}^\dag_{\alpha,m}(t)\hat{a}^{\dag}_{\beta,\ell}(t_1)\rangle_{\rm B}=\langle \hat{a}^\dag_{\alpha,m}(t)\hat{a}^{}_{\beta,n}(t_1)\rangle_{\rm B}=0,
\end{equation}
and
\begin{equation}
\langle \hat{a}_{\alpha,m}(t)\hat{a}^\dag_{\beta,n}(t_1)\rangle_{\rm B}=\delta_{\alpha\beta}\delta_{mn}e^{-i\omega_\alpha (t-t_1)}.
\end{equation}
By averaging over modes $\alpha$, we assume $g_{\alpha,m}$ and $g_{\beta,m}$ are constant over a large enough frequency range, and it leads to
\begin{equation}
\sum_{\alpha,\beta} g_{\alpha,m} g_{\beta, m}\langle \hat{a}_{\alpha,m}(t)\hat{a}^\dag_{\beta,n}(t_1)\rangle_{\rm B}=2\gamma_m\delta_{mn}\delta(t-t_1),
\end{equation}
which is the Markovian approximation.

The effective non-Hermitian hamiltonian reads as
\begin{eqnarray}
\hat{H}_{\rm eff}=\hat{H}_{\rm 0}+\hat{H}_{\rm diss}, \hspace{5ex}
\hat{H}_{\rm diss}&=&\sum_j\left(-i\gamma_m \hat{\cal O}_m^\dag \hat{\cal O}_m^{}+\hat{\cal O}_m^\dag\hat{\xi}_m^{}+\hat{\xi}^\dag_m \hat{\cal O}_m^{}\right),\label{Ham1}
\end{eqnarray}
where $\gamma_m>0$ is dissipation strength and $\hat{\xi}_m(t)$,~$\hat{\xi}_m^\dag(t)$ present the Langevin noise operators. Langevin noises obey the relations
\begin{align}
\label{Eq:xixidag}
&\langle \hat{\xi}_{m}(t)\hat{\xi}_{n}^\dag(t_1)\rangle_{\rm noise}=2\gamma_m\delta_{mn}\delta(t-t_1),\nonumber\\
&\langle \hat{\xi}_m(t)\hat{\xi}_n(t_1)\rangle_{\rm noise}=\langle \hat{\xi}_m^\dag(t)\hat{\xi}_n(t_1)\rangle_{\rm noise}=\langle \hat{\xi}_m^\dag(t)\hat{\xi}^\dag_n(t_1)\rangle_{\rm noise}=0,
\end{align}
where $\langle\cdots\rangle_{\rm noise}$ denotes the noise average.

Now we show that, for any operator $\hat{W}$ in the system,
\begin{equation}
\langle e^{i\hat{H}_{\rm tot}t}\hat{W}e^{-i\hat{H}_{\rm tot}t}\rangle_{\rm B}=\langle e^{i\hat{H}_{\rm eff}t}\hat{W}e^{-i\hat{H}_{\rm eff}t}\rangle_{\rm noise}. \label{equivalence}
\end{equation}

First, we compute the left-hand side of Eq. \eqref{equivalence} order by order in term of $\hat{H}_{\rm int}$. We can define the evolution operator as
\begin{equation}
{\cal U}_{\rm B}(t)\equiv e^{i(\hat{H}_{\rm 0}+\hat{H}_{\rm B})t}e^{-i\hat{H}_{\rm tot}t}=\hat{\cal T}\exp\left(-i\int_0^t \hat{H}_{\rm int}^{\rm I}(t_1)dt_1\right),
\end{equation}
where $\hat{H}_{\rm int}^{\rm I}(t)=e^{i(\hat{H}_{\rm 0}+\hat{H}_{\rm B})t}\hat{H}_{\rm int}e^{-i(\hat{H}_{\rm 0}+\hat{H}_{\rm B})t}$ is in an interaction picture and $\hat{\cal T}$ is the time-ordering operator. Here we also introduce $\tilde{\cal T}$ as anti-time-ordering operator. Then we have
\begin{equation}
\hat{\cal W}_{\rm B}(t)=\langle \hat{\cal U}^\dag_{\rm B}(t)\hat{W}^{\rm I}(t)\hat{\cal U}_{\rm B}(t)\rangle_{\rm B}=\left\langle \tilde{\cal{T}} \exp\left(i\int_0^t \hat{H}_{\rm int}^{\rm I}(t_1)dt_1\right)\hat{W}^{\rm I}(t)\hat{\cal T}\exp\left(-i\int_0^t \hat{H}_{\rm int}^{\rm I}(t_1)dt_1\right)\right\rangle_{\rm B}.
\end{equation}
By perturbation expansion of $\hat{H}_{\rm int}$, we have
\begin{eqnarray}
\hat{\cal W}_{\rm B}(t)&=&\hat{W}^{\rm I}(t)+i\int_0^t \langle \hat{H}_{\rm int}^{\rm I}(t_1)\hat{W}^{\rm I}(t)-\hat{W}^{\rm I}(t)\hat{H}_{\rm int}^{\rm I}(t_1)\rangle_{\rm B}dt_1\nonumber\\
&&-\int_{0\leq t_2\leq t_1\leq t} \!\!\!\!\!\!\!\!\!\!\!\!\! dt_1dt_2 \langle\hat{W}^{I}(t)\hat{H}_{\rm int}^{\rm I}(t_1)\hat{H}_{\rm int}^{\rm I}(t_2)+\hat{V}^{\rm I}(t_2)\hat{H}_{\rm int}^{\rm I}(t_1)\hat{W}^{\rm I}(t)\rangle_{\rm B}+\int_0^t dt_1 dt_2\langle \hat{H}_{\rm int}^{\rm I}(t_1)\hat{W}^{\rm I}(t)\hat{H}_{\rm int}^{\rm I}(t_2)\rangle_{\rm B}+\cdots
\end{eqnarray}
At ${\cal O}(g)$ order, the contribution to $\hat{\cal W}_{\rm B}(t)$ is zero because
\begin{equation}
\langle \hat{H}_{\rm int}^{\rm I}(t_1)\hat{W}^{\rm I}(t)\rangle_{\rm B}=\sum_{m}g_{\alpha, m}^{}\hat{\cal O}_m(t_1)\hat{W}^{\rm I}(t)\langle \hat{a}_{\alpha, m}(t_1)\rangle_{\rm B}=0.
\end{equation}
At ${\cal O}(g^2)$ order, we have
\begin{eqnarray}
&&\int_{0\leq t_2\leq t_1\leq t} \!\!\!\!\!\!\!\!\!\!\!\!\! dt_1dt_2 \langle\hat{W}^{I}(t)\hat{H}_{\rm int}^{\rm I}(t_1)\hat{H}_{\rm int}^{\rm I}(t_2)\rangle_{\rm B}=\int_{0\leq t_2\leq t_1\leq t} \!\!\!\!\!\!\!\!\!\!\!\!\! dt_1dt_2 \sum_m\hat{W}^{\rm I}(t){\cal O}_m^{\rm I, \dag}(t_1)\hat{\cal O}_m^{\rm I}(t_2) \sum_{\alpha,\beta}g_{\alpha, m}g_{\beta, m}\langle \hat{a}_{\alpha, m}^{\rm I}(t_1)\hat{a}^{\rm I, \dag}_{\beta, m}(t_2)\rangle_{\rm B} \nonumber\\
&=&\int_{0\leq t_2\leq t_1\leq t} \!\!\!\!\!\!\!\!\!\!\!\!\! dt_1dt_2 \sum_m\hat{W}^{\rm I}(t){\cal O}_m^{{\rm I},\dag}(t_1)\hat{\cal O}_m^{\rm I}(t_2) 2\gamma_m\delta(t_1-t_2)=\int_0^t dt_1 \sum_m\gamma_m \hat{W}^{\rm I}(t){\cal O}_m^{\rm I,\dag}(t_1)\hat{\cal O}_m^{\rm I}(t_1).
\end{eqnarray}
Here in the first line, we have used the correlation function $\langle \hat{a}^{\rm I}_{\alpha, m}(t)\hat{a}_{\beta, n}^{\rm I}(t_1)\rangle_{\rm B}=\langle \hat{a}^{\rm I,\dag}_{\alpha, m}(t)\hat{a}_{\beta, n}^{\rm I}(t_1)\rangle_{\rm B}=\langle \hat{a}^{\rm I}_{\alpha, m}(t)\hat{a}_{\beta, n}^{\rm I,\dag}(t_1)\rangle_{\rm B}=0$. Similarly, we can compute other terms at ${\cal O}(g^2)$ order, and finally it yields
\begin{eqnarray}
\hat{\cal W}_{\rm B}(t)=\hat{W}^{\rm I}(t)+2\sum_m \gamma_m\int_0^t dt_1\left(\hat{\cal O}^{\rm I,\dag}_m(t_1)\hat{ W}^{\rm I}(t)\hat{\cal O}_m^{\rm I}(t_1)-\frac{1}{2}\left\{\hat{W}^{\rm I}(t),\hat{\cal O}^{\rm I,\dag}_m(t_1)\hat{\cal O}_m^{\rm I}(t_1)\right\}\right)+\cdots.\label{Env}
\end{eqnarray}

Second, we compute the right-hand side of Eq. \eqref{equivalence} order by order in term of $\hat{H}_\text{diss}$. We can define the evolution operator as
\begin{equation}
\hat{\cal U}(t)=e^{i \hat{H}_{\rm 0}t}e^{-i\hat{H}_{\rm eff}t}=\hat{\cal T}\exp\left(-i\int_0^t \hat{H}^{\rm I}_{\rm diss}(t_1)dt_1\right),\label{TimeU}
\end{equation}
and we have
\begin{equation}
\hat{\cal W}_{\rm B}(t)=\langle \hat{\cal U}^\dag(t)\hat{W}^{\rm I}(t)\hat{\cal U}(t)\rangle_{\rm noise}=\left\langle \tilde{\cal{T}} \exp\left(i\int_0^t \hat{H}_{\rm diss}^{\rm I,\dag}(t_1)dt_1\right)\hat{W}^{\rm I}(t)\hat{\cal T}\exp\left(-i\int_0^t \hat{H}_{\rm diss}^{\rm I}(t_1)dt_1\right)\right\rangle_{\rm noise}.
\end{equation}
Now we expand $\hat{\cal W}(t)$ to $\gamma$'s first order and $\xi$'s second order, we have
\begin{eqnarray}
\hat{\cal W}(t)&=&\hat{W}^{\rm I}(t)+i\int_0^t \langle \hat{H}_{\rm diss}^{\rm I,\dag}(t_1)\hat{W}^{\rm I}(t)-\hat{W}^{\rm I}(t)\hat{H}_{\rm diss}^{\rm I}(t_1)\rangle_{\rm noise}dt_1+\int_0^t dt_1 dt_2\langle \hat{H}_{\rm diss}^{\rm I}(t_1)\hat{W}^{\rm I}(t)\hat{H}_{\rm diss}^{\rm I}(t_2)\rangle_{\rm noise}+\cdots.
\end{eqnarray}
Here we can find that
\begin{equation}
i\langle \hat{H}_{\rm diss}^{\rm I,\dag}(t_1)\hat{W}^{\rm I}(t)-\hat{W}^{\rm I}(t)\hat{H}_{\rm diss}^{\rm I}(t_1)\rangle_{\rm noise}=-\sum_m\gamma_m\left\{\hat{W}^{\rm I}(t),\hat{\cal O}_m^{\rm I,\dag}(t_1)\hat{\cal O}_m^{\rm I}(t_1)\right\},\label{noise01}
\end{equation}
where we used $\langle \hat{\xi}(t)\rangle_{\rm noise}=0$, and
\begin{eqnarray}
&&\int_0^t dt_1 dt_2\langle \hat{H}_{\rm diss}^{\rm I}(t_1)\hat{W}^{\rm I}(t)\hat{H}_{\rm diss}^{\rm I}(t_2)\rangle_{\rm noise}=\sum_m \int_0^t dt_1 dt_2 \langle \hat{\xi}^{}_m(t_1)\hat{\cal O}_m^{\rm I,\dag}(t_1)\hat{W}^{\rm I}(t)\hat{\xi}_m^\dag(t_2)\hat{\cal O}^{\rm I}_m(t_2)\rangle_{\rm noise}\nonumber\\
&=&2\sum_m\gamma_m\int_0^t dt_1 dt_2 \delta(t_1-t_2)\hat{\cal O}_m^{\rm I,\dag}(t_1)\hat{W}^{\rm I}(t)\hat{\cal O}^{\rm I}_m(t_2).\label{noise02}
\end{eqnarray}
Combining Eq.~\eqref{noise01} and Eq.~\eqref{noise02} , we have
\begin{eqnarray}
\hat{\cal W}(t)=\hat{W}^{\rm I}(t)+2\sum_m \gamma_m\int_0^t dt_1\left(\hat{\cal O}^{\rm I,\dag}_m(t_1)\hat{ W}^{\rm I}(t)\hat{\cal O}_m^{\rm I}(t_1)-\frac{1}{2}\left\{\hat{W}^{\rm I}(t),\hat{\cal O}^{\rm I,\dag}_m(t_1)\hat{\cal O}_m^{\rm I}(t_1)\right\}\right)+\cdots.\label{Noise}
\end{eqnarray}

Comparing Eq.~\eqref{Env} with Eq.~\eqref{Noise}, we have shown that Eq. \eqref{equivalence} is valid up to $\gamma$'s first order. These two expressions can be further calculated by higher order expansion. Since we have assumed that the bath is consist of free bosonic modes, there is no irreducible four-points or high-points Green's functions, therefore all these correlation functions emerged in the higher order expansion can be factorized into two-points Green's function $\sum_{\alpha,\beta}g_{\alpha, m}g_{\beta, m}\langle \hat{a}_{\alpha, m}^{\rm I}(t_1)\hat{a}^{\rm I, \dag}_{\beta, m}(t_2)\rangle_{\rm B}$. Hence, we can order by order shown that Eq. \eqref{equivalence} is valid up to all orders.

\section{II. The Full Expression of Green's Function} \label{subsec22}
Here we show the full expression of the Green's function discussed in the main-text. The retarded Green function is defined by
\begin{align}
\mathcal{G}_{ij}\left(t,0\right)=-i\Theta(t)\left\langle \left\{ \hat{c}_{i}(t),\hat{c}_{j}^{\dagger}\left(0\right)\right\} \right\rangle,
\end{align}
where $\Theta(t)$ is the Heaviside step function, which is unit for $t\geq 0$ and zero for $t<0$.
The average state is taken as the Fermi sea $|\rm FS \rangle$. And the Fermi energy is taken as $E_F=0$.
$\hat{c}_{i}$ ($\hat{c}_{j}^{\dagger}$) denotes the annihilation (creation) operator corresponding to the eigenstate $\ket{\Psi_i}$ ($\ket{\Psi_j}$) and $\hat{c}_{i}(t)=e^{i \hat{H}_{\rm eff} t} \hat{c}_{i} e^{-i \hat{H}_{\rm eff} t}$.
By introducing the greater Green's function $\mathcal{\mathcal{G}}_{ij}^{>}(t,0)=-i\langle c_{i}(t)c_{j}^{\dagger}(0)\rangle$ and the lesser Green's function $\mathcal{\mathcal{G}}_{ij}^{<}(t,0)=i\langle c_{j}^{\dagger}(0)c_{i}(t)\rangle$, the retarded Green's function can be rewritten as ${\mathcal{G}}_{ij}(t,0)=\Theta(t)[\mathcal{G}_{ij}^{>}(t,0)-\mathcal{G}_{ij}^{<}(t,0)]$. In the interaction picture, the greater and lesser Green's function can be expressed as
\begin{align}
\mathcal{\mathcal{G}}_{ij}^{>}(t,0)&=-i\left\langle \hat{\cal U}^{\dagger}(t)\hat{c}_{i}^{\rm I}(t)\hat{\cal U}(t)\hat{c}_{j}^{\rm I,\dagger}(0)\right\rangle,  \label{Ggreater}\\
\mathcal{\mathcal{G}}_{ij}^{<}(t,0)&=i\left\langle\hat{c}_{j}^{\rm I,\dagger}(0)\hat{\cal U}^{\dagger}(t)\hat{c}_{i}^{\rm I}(t)\hat{\cal U}(t)\right\rangle.  \label{Glesser}
\end{align}
Inserting Eq.~\eqref{TimeU} into Eq.~\eqref{Ggreater} and Eq.~\eqref{Glesser}, then taking the noise average and keeping the first-order of $\gamma_m$,  we have
\begin{align}
\mathcal{G}_{ij}^{>}\left(t,0\right) & =-i\left\langle \hat{c}_{i}^{\rm I}(t)\hat{c}_{j}^{\rm I,\dagger}\left(0\right)\right\rangle
+i\sum_{m}\gamma_m\int_{0}^{t}dt_{1}\left\langle \hat{c}_{i}^{\rm I}(t)\hat{\mathcal{O}}_m^{\rm I,\dagger}\left(t_{1}\right)\hat{\mathcal{O}}_m^{\rm I}\left(t_{1}\right)\hat{c}_{j}^{\rm I,\dagger}\left(0\right)\right\rangle\nonumber\\
&+i\sum_{m}\gamma_m\int_{0}^{t}dt_{1}\left\langle \hat{\mathcal{O}}_m^{\rm I,\dagger}\left(t_{1}\right)\hat{\mathcal{O}}_m^{\rm I}\left(t_{1}\right)\hat{c}_{i}^{\rm I}(t)\hat{c}_{j}^{\rm I,\dagger}\left(0\right)\right\rangle
-2i\sum_{m}\gamma_m\int_{0}^{t}dt_{1}\left\langle \hat{\mathcal{O}}_m^{\rm I,\dagger}\left(t_{1}\right)\hat{c}_{i}^{\rm I}(t)\hat{\mathcal{O}}_m^{\rm I}\left(t_{1}\right)\hat{c}_{j}^{\rm I,\dagger}\left(0\right)\right\rangle,
\label{Ggreater1}
\end{align}
and
\begin{align}
\mathcal{G}_{ij}^{<}\left(t,0\right)
&=i\left\langle \hat{c}_{j}^{\rm I,\dagger}\left(0\right)\hat{c}_{i}^{\rm I}(t)\right\rangle
-i\sum_{m}\gamma_m\int_{0}^{t}dt_{1}\left\langle \hat{c}_{j}^{\rm I,\dagger}\left(0\right)\hat{\mathcal{O}}_m^{\rm I,\dagger}\left(t_{1}\right)\hat{\mathcal{O}}_m^{\rm I}\left(t_{1}\right)\hat{c}_{i}^{\rm I}(t)\right\rangle \nonumber\\
&-i\sum_{m}\gamma_m\int_{0}^{t}dt_{1}\left\langle \hat{c}_{j}^{\rm I,\dagger}\left(0\right)\hat{c}_{i}^{\rm I}(t)\hat{\mathcal{O}}_m^{\rm I,\dagger}\left(t_{1}\right)\hat{\mathcal{O}}_m^{\rm I}\left(t_{1}\right)\right\rangle
+2i\sum_{m}\gamma_m\int_{0}^{t}\left\langle \hat{c}_{j}^{\rm I,\dagger}\left(0\right)\hat{\mathcal{O}}_m^{\rm I,\dagger}\left(t_{1}\right)\hat{c}_{i}^{\rm I}(t)\hat{\mathcal{O}}_m^{\rm I}\left(t_{1}\right)\right\rangle.
\label{Glesser1}
\end{align}
Hence, the the full expressions for zero- and first-order of dissipation strength are written by
\begin{align}
\mathcal{G}_{i j}^{(0)}=&-i \Theta(t)\left\langle\left\{\hat{c}_{i}^{\mathrm{I}}(t), \hat{c}_{j}^{\dagger, \mathrm{I}}(0)\right\}\right\rangle\nonumber \\
\mathcal{G}_{i j}^{(1)}= &i\sum_{m}\gamma_m\int_{0}^{t}dt_{1}\left\langle \hat{c}_{i}^{\rm I}(t)\hat{\mathcal{O}}_m^{\rm I,\dagger}\left(t_{1}\right)\hat{\mathcal{O}}_m^{\rm I}\left(t_{1}\right)\hat{c}_{j}^{\rm I,\dagger}\left(0\right)\right\rangle
+i\sum_{m}\gamma_m\int_{0}^{t}dt_{1}\left\langle \hat{\mathcal{O}}_m^{\rm I,\dagger}\left(t_{1}\right)\hat{\mathcal{O}}_m^{\rm I}\left(t_{1}\right)\hat{c}_{i}^{\rm I}(t)\hat{c}_{j}^{\rm I,\dagger}\left(0\right)\right\rangle \nonumber \\
&-2i\sum_{m}\gamma_m\int_{0}^{t}dt_{1}\left\langle \hat{\mathcal{O}}_m^{\rm I,\dagger}\left(t_{1}\right)\hat{c}_{i}^{\rm I}(t)\hat{\mathcal{O}}_m^{\rm I}\left(t_{1}\right)\hat{c}_{j}^{\rm I,\dagger}\left(0\right)\right\rangle
-i\sum_{m}\gamma_m\int_{0}^{t}dt_{1}\left\langle \hat{c}_{j}^{\rm I,\dagger}\left(0\right)\hat{\mathcal{O}}_m^{\rm I,\dagger}\left(t_{1}\right)\hat{\mathcal{O}}_m^{\rm I}\left(t_{1}\right)\hat{c}_{i}^{\rm I}(t)\right\rangle \nonumber\\
&-i\sum_{m}\gamma_m\int_{0}^{t}dt_{1}\left\langle \hat{c}_{j}^{\rm I,\dagger}\left(0\right)\hat{c}_{i}^{\rm I}(t)\hat{\mathcal{O}}_m^{\rm I,\dagger}\left(t_{1}\right)\hat{\mathcal{O}}_m^{\rm I}\left(t_{1}\right)\right\rangle
+2i\sum_{m}\gamma_m\int_{0}^{t}\left\langle \hat{c}_{j}^{\rm I,\dagger}\left(0\right)\hat{\mathcal{O}}_m^{\rm I,\dagger}\left(t_{1}\right)\hat{c}_{i}^{\rm I}(t)\hat{\mathcal{O}}_m^{\rm I}\left(t_{1}\right)\right\rangle.
\end{align}
In order to see the break of degeneracy, we project the operators $\hat{\mathcal{O}}^{\rm I}_m$, $\hat{\mathcal{O}}_m^{\rm I,\dagger}$ and $\hat{\mathcal{O}}_m^{\rm I,\dagger}\hat{\mathcal{O}}_m^{\rm I}$ onto Kramers space and define $\mathcal{G}_{i j,\rm K}^{(1)}$ as
\begin{align}
\mathcal{G}_{i j,\rm K}^{(1)}= &i\sum_{m}\gamma_m\Bigg\{\int_{0}^{t}dt_{1}\left\langle \hat{c}_{i}^{\rm I}(t)\left[\hat{\Pi}_\text{K}\hat{\mathcal{O}}_m^{\rm I,\dagger}\left(t_{1}\right)\hat{\mathcal{O}}_m^{\rm I}\left(t_{1}\right)\hat{\Pi}_\text{K}\right]\hat{c}_{j}^{\rm I,\dagger}\left(0\right)\right\rangle
+\int_{0}^{t}dt_{1}\left\langle \left[\hat{\Pi}_\text{K}\hat{\mathcal{O}}_m^{\rm I,\dagger}\left(t_{1}\right)\hat{\mathcal{O}}_m^{\rm I}\left(t_{1}\right)\hat{\Pi}_\text{K}\right]\hat{c}_{i}^{\rm I}(t)\hat{c}_{j}^{\rm I,\dagger}\left(0\right)\right\rangle \nonumber \\
&-2\int_{0}^{t}dt_{1}\left\langle \left[\hat{\Pi}_\text{K}\hat{\mathcal{O}}_m^{\rm I,\dagger}\left(t_{1}\right)\hat{\Pi}_\text{K}\right]\hat{c}_{i}^{\rm I}(t)\left[\hat{\Pi}_\text{K}\hat{\mathcal{O}}_m^{\rm I}\left(t_{1}\right)\hat{\Pi}_\text{K}\right]\hat{c}_{j}^{\rm I,\dagger}\left(0\right)\right\rangle
-\int_{0}^{t}dt_{1}\left\langle \hat{c}_{j}^{\rm I,\dagger}\left(0\right)\hat{c}_{i}^{\rm I}(t)\left[\hat{\Pi}_\text{K}\hat{\mathcal{O}}_m^{\rm I,\dagger}\left(t_{1}\right)\hat{\mathcal{O}}_m^{\rm I}\left(t_{1}\right)\hat{\Pi}_\text{K}\right]\right\rangle   \nonumber\\
&-\int_{0}^{t}dt_{1}\left\langle \hat{c}_{j}^{\rm I,\dagger}\left(0\right)\left[\hat{\Pi}_\text{K}\hat{\mathcal{O}}_m^{\rm I,\dagger}\left(t_{1}\right)\hat{\mathcal{O}}_m^{\rm I}\left(t_{1}\right)\hat{\Pi}_\text{K}\right]\hat{c}_{i}^{\rm I}(t)\right\rangle
+2\int_{0}^{t}\left\langle \hat{c}_{j}^{\rm I,\dagger}\left(0\right)\left[\hat{\Pi}_\text{K}\hat{\mathcal{O}}_m^{\rm I,\dagger}\left(t_{1}\right)\hat{\Pi}_\text{K}\right]\hat{c}_{i}^{\rm I}(t)\left[\hat{\Pi}_\text{K}\hat{\mathcal{O}}_m^{\rm I}\left(t_{1}\right)\hat{\Pi}_\text{K}\right]\right\rangle\Bigg\}.
\end{align}
The following relations can be deduced by means of Schur's Lemma as long as $\hat{\mathcal{O}}^{\dagger}$ is non-Hermitian operator
\begin{align}
\hat{\Pi}_{K} \hat{\mathcal{O}}^{\dagger, \mathrm{I}}(t_1)\hat{\mathcal{O}}^{\mathrm{I}}(t_1)  \hat{\Pi}_{\mathrm{K}}=\hat{\Pi}_{\mathrm{K}}  \hat{\mathcal{O}}^{\dagger}\hat{\mathcal{O}} \hat{\Pi}_{\mathrm{K}}\propto \hat{\rm I}, \nonumber\\
\hat{\Pi}_{K} \hat{\mathcal{O}}^{\dagger, \mathrm{I}}(t_1) \hat{\Pi}_{\mathrm{K}}=\hat{\Pi}_{\mathrm{K}}  \hat{\mathcal{O}}^{\dagger} \hat{\Pi}_{\mathrm{K}}\not\propto \hat{\rm I}, \nonumber\\
\hat{\Pi}_{K} \hat{\mathcal{O}}^{\mathrm{I}}(t_1)  \hat{\Pi}_{\mathrm{K}}=\hat{\Pi}_{\mathrm{K}}  \hat{\mathcal{O}} \hat{\Pi}_{\mathrm{K}}\not\propto \hat{\rm I}.
\end{align}
which is directly responsible to break of degeneracy as discussed in maintext.

\section{III. The Full Expression of Impurity Matrix Element} \label{subsec23}
Here we discuss the  matrix element of a local impurity potential between two Kramers states
\begin{align}
\label{Eq:Tijt}
V_{ij}(t)=\bra{\Psi_i}\hat{V}(t)\ket{\Psi_j},
\end{align}
and the time-dependent impurity potential $\hat{V}(t)$ is
\begin{align}
\label{VHeis}
\hat{V}(t)=e^{i\hat{H}_{{\rm eff}}t}\hat{V}e^{-i\hat{H}_{{\rm eff}}t}=\hat{\cal U}^{\dagger}(t)\hat{V}^{\rm I}(t)\hat{\cal U}(t).
\end{align}
To the first order of $\gamma$, Eq. \eqref{VHeis} can be expanded as
\begin{align}
\label{Eq:Eijt2}
V_{ij}(t)&=\bra{\Psi_i}\hat{V}^{\rm I}(t)\ket{\Psi_j}-\int_{0}^{t}dt_{1}\sum_{m}\gamma_{m}\Big[\bra{\Psi_i}\hat{\mathcal{O}}_{m}^{\rm I,\dagger}(t_{1})\hat{\mathcal{O}}_{m}^{\rm I}(t_{1})\hat{V}^{\rm I}(t)\ket{\Psi_j}\nonumber\\
&+\bra{\Psi_i}\hat{V}^{\rm I}(t)\hat{\mathcal{O}}_{m}^{\rm I,\dagger}(t_{1})\hat{\mathcal{O}}_{m}^{\rm I}(t_{1})\ket{\Psi_j}-2\bra{\Psi_i}\hat{\mathcal{O}}_{m}^{\rm I,\dagger}(t_{1})\hat{V}^{\rm I}(t)\hat{\mathcal{O}}_{m}^{\rm I}(t_{1})\ket{\Psi_j}\Big].
\end{align}
Since $\ket{\Psi_i}$, $\ket{\Psi_j}$ are eigenstates belonging to the Kramers degenerate space, one can see that $\bra{\Psi_i}\hat{V}^{\rm I}(t)\ket{\Psi_j}=V_0\delta_{ij}$ which  $V_0$ is proportional to impurity strength and Eq. \eqref{Eq:Eijt2} can be written as
\begin{align}
\label{Eq:Eijt3}
V_{ij}(t)&=V_0\delta_{ij}-\int_{0}^{t}dt_{1}\sum_{m}\gamma_{m}\Big[\bra{\Psi_i}\hat{\Pi}_{\rm K}\hat{\mathcal{O}}_{m}^{\rm I,\dagger}(t_{1})\hat{\mathcal{O}}_{m}^{\rm I}(t_{1})\hat{V}^{\rm I}(t)\hat{\Pi}_{\rm K}\ket{\Psi_j}\nonumber\\
&+\bra{\Psi_i}\hat{\Pi}_{\rm K}\hat{V}^{\rm I}(t)\hat{\mathcal{O}}_{m}^{\rm I,\dagger}(t_{1})\hat{\mathcal{O}}_{m}^{\rm I}(t_{1})\hat{\Pi}_{\rm K}\ket{\Psi_j}-2\bra{\Psi_i}\hat{\Pi}_{\rm K}\hat{\mathcal{O}}_{m}^{\rm I,\dagger}(t_{1})\hat{V}^{\rm I}(t)\hat{\mathcal{O}}_{m}^{\rm I}(t_{1})\hat{\Pi}_{\rm K}\ket{\Psi_j}\Big].
\end{align}
This is the full expression of matrix element $V_{ij}$ to the first-order correction of $\gamma_m$.

\section{IV. Detailed Calculations in Kane-Mele Model with Dissipation} \label{sec2}

In this section, we'll give the detailed set-up of the Kane-Mele model and calculations of Fig.1 in the main text.
\subsection{A. The detailed set-up of Kane-Mele model}
\begin{figure}[t]
	\centering
	\includegraphics[width=0.8\textwidth]{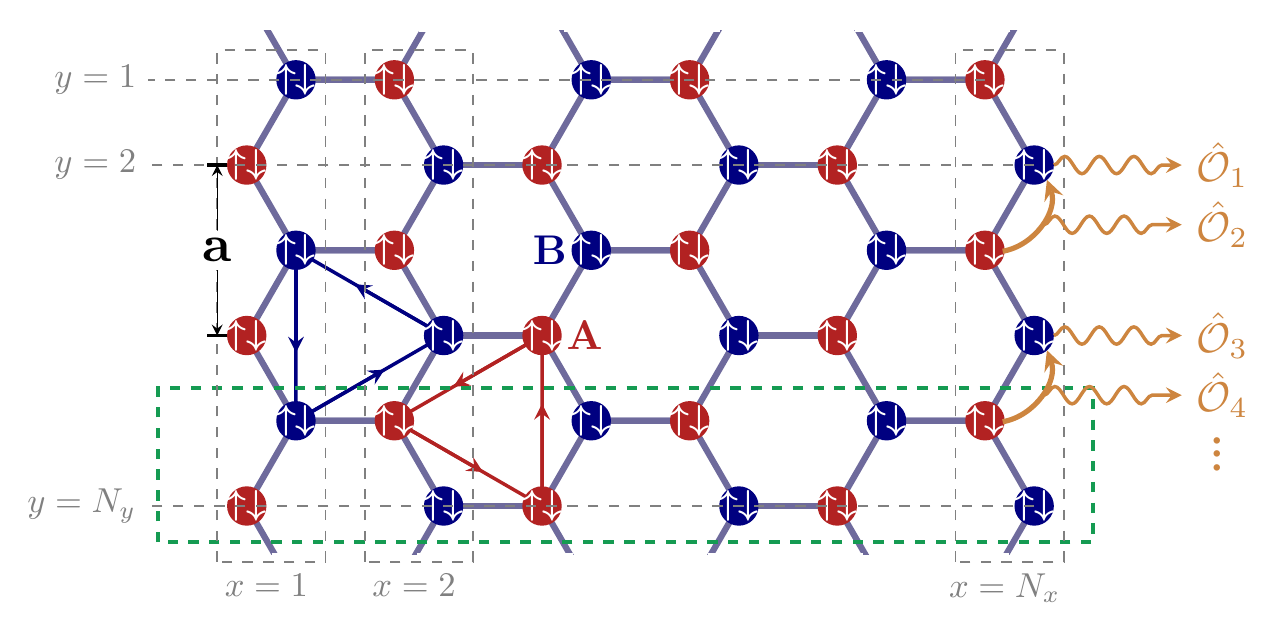}
	\caption{
		Honeycomb lattice of a finite-sized zigzag ribbon in the tight-binding Kane-Mele model with open boundary condition along $x$-axis and with periodical boundary condition along $y$-axis.  The sample size is $N_x\times N_y$. The green-dashed rectangle represents for the super-unit-cell of the zigzag ribbon,
		which repeats itself along $y$-axis. The coupling operators  $\hat{\mathcal{O}}_m$ are located on the right edge, which are defined on links for odd m and defined on sites for even m.
	}
	\label{KanemeleSupp}
\end{figure}

\begin{figure}[t]
	\centering
	\includegraphics[width=0.6\textwidth]{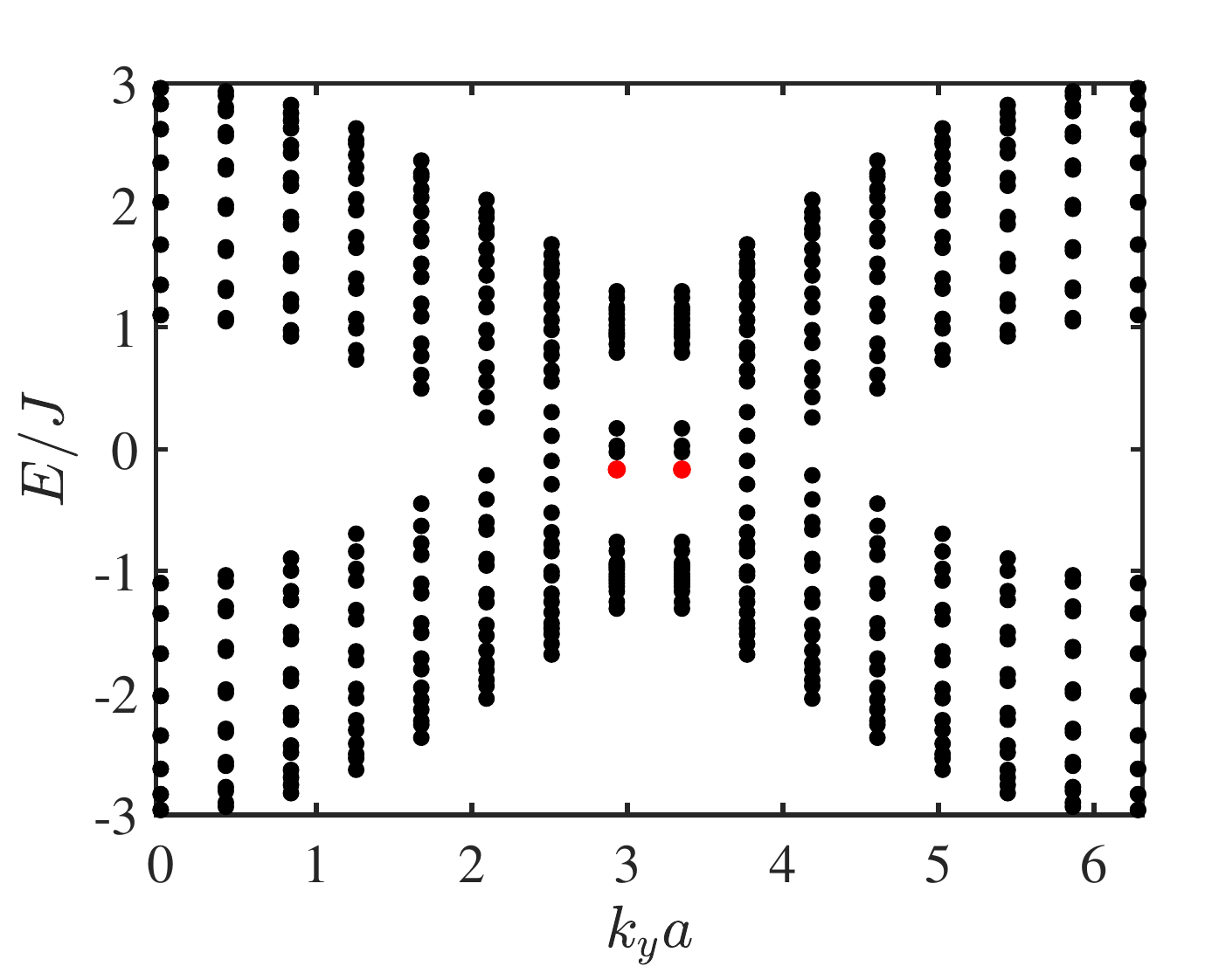}
	\caption{
		Eigen-energy of the Kane-Mele Hamiltonian $\hat{H}_0$ as a function of $k_y$. The red dots represent for a pair of Kramers degenerate states considered in the maintext. Here the parameters are the same ones configured in Fig. 1 of the maintext.
	}
	\label{KanemeleSupp2}
\end{figure}

We start from the following effective non-Hermitian Hamiltonian
\begin{eqnarray}
\label{NHHeff}
\hat{H}=\hat{H}_0+\hat{H}_{\rm diss}. \label{Ham_S1}
\end{eqnarray}
For the $\hat{H}_0$ part, we take it as the Kane-Mele Hamiltonian
\begin{eqnarray}
\hat{H}_{0}=J\sum_{\langle i,j\rangle,s}\hat{c}_{i,s}^{\dagger}\hat{c}_{j,s}+i\lambda_{\text{SO}}\sum_{\langle\langle i,j\rangle\rangle,s,s^\prime}\nu_{ij}\hat{c}_{i,s}^{\dagger}\sigma_{ss^{\prime}}^{z}\hat{c}_{j,s^\prime}+i\lambda_{\text{R}}\sum_{\langle i,j\rangle,s,s'}\hat{c}_{i,s}^{\dagger}({\bm{\sigma}}\times{\bm{d}}_{ij})_{ss^{\prime}}^{z}\hat{c}_{j,s^{\prime}}+\lambda_{\nu}\sum_{i,s}\xi_{i}\hat{c}_{i,s}^{\dagger}\hat{c}_{i,s},
\end{eqnarray}
where $\nu_{ij}$ is $1$ ($-1$) when the coupling direction from $j$ to $i$ is (not) consistent with the blue or red arrow direction shown in Fig.~\ref{KanemeleSupp}, $\bm{d}_{ij}$ is the unit vector from $i$ to $j$, and $\xi_i$ is  $1$ ($-1$) when $i$ site is on the sublattice $A$ ($B$).
For the Kane-Mele lattice model, we apply a zigzag boundary condition and define a super unit cell as shown in Fig.~\ref{KanemeleSupp} . In such configuration, quasi-momentum $k_y$ is a good quantum number. Therefore we could diagnolize $\hat{H}_0$ for different $k_y$,and plot eigenenergy in Fig.~\ref{KanemeleSupp2}. The red dot represents for the Kramers states we have considered in the calculation of maintext.

The dissipative part is given by
\begin{eqnarray}
\hat{H}_{{\rm diss}}=\sum_{m}\left(-i\gamma_{m}\hat{{\cal O}}_{m}^{\dag}\hat{{\cal O}}_{m}+\hat{{\cal O}}_{m}^{\dag}\xi_{m}+\xi_{m}^{\dag}\hat{{\cal O}}_{m}\right).
\label{KM}
\end{eqnarray}
The dissipation coupling operator $\hat{\mathcal{O}}_m$ is located on the zigzag edge of the honeycomb lattice as shown in Fig.~\ref{KanemeleSupp}. When $m$ is even number, $\hat{\mathcal{O}}_m$ is defined  on the link $\langle i_m, j_m\rangle$
\begin{eqnarray}
\label{Om1}
\hat{\mathcal{O}}_m=\sum_{s} c_{i_m, s}^{\dagger} c_{j_m, s},
\end{eqnarray}
where the site $i_m$ represents for sublattice B with location $x_{i_m}=N_x, y_{i_m}=m$ and the site $j_m$ represents for sublattice A with location $x_{j_m}=N_x, y_{j_m}=m+1$.
When $m$ is odd number, $\hat{\mathcal{O}}_m$ is defined  on the site $i_m$
\begin{eqnarray}
\label{Om2}
\hat{\mathcal{O}}_m=\sum_{s, s^{\prime}} i c_{i_m, s}^{\dagger} \sigma_{s s^{\prime}}^{y} c_{i_m, s^{\prime}},
\end{eqnarray}
where the site $i_m$ represents for sublattice B with location $x_{ i_m}=N_x, y_{i_m}=m+1$.

In Fig. 1(c), we calculate the matrix element of a local impurity potential. Here, we give the impurity potential as
\begin{eqnarray}
\hat{V}=V\sum_s \hat{c}_{i,s}^{\dagger} \hat{c}_{i,s},
\end{eqnarray}
where $V$ is the impurity strength and $i$ is the impurity position. In the calculation of Fig. 1(c), $i$ is taken as $x_i=N_x$ and $y_i=2$.

\subsection{B. Calculation of Green's function in Kane-Mele model}
In this subsection, we'll give the detailed  calculations of Green's function given in section \ref{subsec22}.
Starting from Eq.~(\ref{Ggreater1}) and Eq.~(\ref{Glesser1}), since we have considered $\hat{\mathcal{O}}_{m}$ is quadratic fermion operator as Eq.~(\ref{Om1}) and Eq.~(\ref{Om2}), $\hat{\mathcal{O}}_{m}$ can be expressed  in the second quantization form as
\begin{align}
\hat{\mathcal{O}}_{m}^{\dagger}=\sum_{\ell_{1},\ell_{2}}\bra{\Psi_{\ell_{1}}}\hat{\mathcal{O}}_{m}^{\dagger}\ket{\Psi_{\ell_{2}}}\hat{c}_{\ell_{1}}^{\dagger}c_{\ell_{2}},\\
\hat{\mathcal{O}}_{m}=\sum_{\ell_{3},\ell_{4}}\bra{\Psi_{\ell_{3}}}\hat{\mathcal{O}}_{m}\ket{\Psi_{\ell_{4}}}\hat{c}_{\ell_{3}}^{\dagger}c_{\ell_{4}},
\end{align}
where $\ket{\Psi_{\ell_{n}}}$ is the eigen-state of $\hat{H}_0$ with the eigen energy $E_{\ell_{n}}$. Then Eq.~(\ref{Ggreater1}) and Eq.~(\ref{Glesser1}) can be rewritten as

\begin{align}
\mathcal{G}_{ij}^{>}\left(t,0\right)&=-i\left\langle \hat{c}_{i}^{{\rm I}}(t)\hat{c}_{j}^{{\rm I,\dagger}}\left(0\right)\right\rangle +\sum_{\ell_{1},\ell_2,\ell_3,\ell_4}(-i\widetilde{M}_{\ell_{1},\ell_{2},\ell_{3},\ell_{4}})\left[-\int_{0}^{t}dt_{1}\left\langle \hat{c}_{i}^{{\rm I}}(t)\hat{c}_{\ell_{1}}^{{\rm I,\dagger}}(t_{1})\hat{c}_{\ell_{2}}^{{\rm I}}(t_{1})\hat{c}_{\ell_{3}}^{{\rm I,\dagger}}(t_{1})\hat{c}_{\ell_{4}}^{{\rm I}}\left(t_{1}\right)\hat{c}_{j}^{{\rm I,\dagger}}\left(0\right)\right\rangle \right.\nonumber\\
&-\int_{0}^{t}dt_{1}\left\langle \hat{c}_{\ell_{1}}^{{\rm I,\dagger}}(t_{1})\hat{c}_{\ell_{2}}^{{\rm I}}(t_{1})\hat{c}_{\ell_{3}}^{{\rm I,\dagger}}(t_{1})\hat{c}_{\ell_{4}}^{{\rm I}}\left(t_{1}\right)\hat{c}_{i}^{{\rm I}}(t)\hat{c}_{j}^{{\rm I,\dagger}}\left(0\right)\right\rangle +2\int_{0}^{t}dt_{1}\left\langle \hat{c}_{\ell_{1}}^{{\rm I,\dagger}}(t_{1})\hat{c}_{\ell_{2}}^{{\rm I}}\left(t_{1}\right)\hat{c}_{i}^{{\rm I}}(t)\hat{c}_{\ell_{3}}^{{\rm I,\dagger}}(t_1)\hat{c}_{\ell_{4}}^{{\rm I}}\left(t_{1}\right)\hat{c}_{j}^{{\rm I,\dagger}}\left(0\right)\right\rangle,
\label{Ggreater3}
\end{align}
and
\begin{align}
\label{Glesser3}
\mathcal{G}_{ij}^{<}\left(t,0\right)=i\left\langle \hat{c}_{j}^{{\rm I,\dagger}}\left(0\right)\hat{c}_{i}^{{\rm I}}(t)\right\rangle +\sum_{\ell_{1},\ell_2,\ell_3,\ell_4}i\widetilde{M}_{\ell_{1},\ell_{2},\ell_{3},\ell_{4}}\left[-\int_{0}^{t}dt_{1}\left\langle \hat{c}_{j}^{{\rm I,\dagger}}\left(0\right)\hat{c}_{\ell_{1}}^{{\rm I},\dagger}(t_{1})\hat{c}_{\ell_{2}}^{{\rm I}}(t_{1})\hat{c}_{\ell_{3}}^{{\rm I},\dagger}(t_{1})\hat{c}_{\ell_{4}}^{{\rm I}}(t_{1})\hat{c}_{i}^{{\rm I}}(t)\right\rangle \right.\nonumber
\\-\int_{0}^{t}\left\langle \hat{c}_{j}^{{\rm I,\dagger}}\left(0\right)\hat{c}_{i}^{{\rm I}}(t)\hat{c}_{\ell_{1}}^{{\rm I},\dagger}(t_{1})\hat{c}_{\ell_{2}}^{{\rm I}}(t_{1})\hat{c}_{\ell_{3}}^{{\rm I},\dagger}(t_{1})\hat{c}_{\ell_{4}}^{{\rm I}}\left(t_{1}\right)\right\rangle dt_{1}\left.+2\int_{0}^{t}dt_{1}\left\langle \hat{c}_{j}^{{\rm I,\dagger}}\left(0\right)\hat{c}_{\ell_{1}}^{{\rm I,\dagger}}(t_{1})\hat{c}_{\ell_{2}}^{{\rm I}}(t_{1})\hat{c}_{i}^{{\rm I}}(t)\hat{c}_{\ell_{3}}^{{\rm I,\dagger}}(t_{1})\hat{c}_{\ell_{4}}^{{\rm I}}\left(t_{1}\right)\right\rangle \right],
\end{align}
by defining $\widetilde{M}_{\ell_{1},\ell_{2},\ell_{3},\ell_{4}}$ as
\begin{align}
\widetilde{M}_{\ell_{1},\ell_{2},\ell_{3},\ell_{4}}=\sum_{m}\gamma_{m}\bra{\Psi_{\ell_{1}}}\hat{\mathcal{O}}_{m}^{\dagger}\ket{\Psi_{\ell_{2}}}\bra{\Psi_{\ell_{3}}}\hat{\mathcal{O}}_{m}\ket{\Psi_{\ell_{4}}}.
\end{align}
Now we apply the Wick's theorem in above formula. Firstly, we should note the contribution from $\left\langle \hat{c}_{i}^{\rm I}(t)\hat{c}_{j}^{\rm I,\dagger}(0)\right\rangle \left\langle \hat{c}_{\ell_{1}}^{\rm I,\dagger}\hat{c}_{\ell_{2}}^{\rm I}\hat{c}_{\ell_{3}}^{\rm I,\dagger}\hat{c}_{\ell_{4}}^{\rm I}(t_{1})\right\rangle$ in $\mathcal{G}_{ij}^{>}(t,0)$ is canceled out by summing all the integrations. In the same way, the contribution of $\left\langle \hat{c}_{j}^{{\rm I,\dagger}}\left(0\right)\hat{c}_{i}^{{\rm I}}(t)\right\rangle \left\langle \hat{c}_{\ell_{1}}^{\rm I,\dagger}\hat{c}_{\ell_{2}}^{\rm I}\hat{c}_{\ell_{3}}^{\rm I,\dagger}\hat{c}_{\ell_{4}}^{\rm I}(t_{1})\right\rangle $ in $\mathcal{G}_{ij}^{<}(t,0)$ also vanishes. Then we consider  the Green's function of a pair of Kramers degenerate states inside the Fermi sea, which leads to $E_i=E_{j}=E_0$, $n_i=n_{j}=1$ and two point correlation functions $\langle  {\rm {FS}}|...\hat{c}_{i(j)}^{\rm I,\dagger}(t_{1})|{ \rm FS}\rangle=\langle{ \rm FS}|\hat{c}_{i(j)}^{\rm I}(t_{1})...|{\rm FS}\rangle=0$. Therefore, Eq.~\eqref{Ggreater3} and Eq.~\eqref{Glesser3} is reduced to
\begin{align}
\mathcal{G}_{ij}^{>}(t,0)&=0,\\
\mathcal{G}_{ij}^{<}(t,0)&=i\langle\hat{c}_{j}^{\rm I,\dagger}(0)\hat{c}_{i}^{\rm I}(t)\rangle
+\sum_{l_{1}}\sum_{\ell_{2}}\sum_{\ell_{3}}\sum_{\ell_{4}}i\widetilde{M}_{l_{1},\ell_{2},\ell_{3},\ell_{4}}\nonumber\\
&\times\Big\{\int_{0}^{t}dt_{1}\big[\langle\hat{c}_{j}^{\rm I,\dagger}(0)\hat{c}_{\ell_{4}}^{\rm I}(t_{1})\rangle\langle\hat{c}_{\ell_{3}}^{\rm I,\dagger}(t_{1})\hat{c}_{i}^{\rm I}(t)\rangle\langle\hat{c}_{l_{1}}^{\rm I,\dagger}(t_{1})\hat{c}_{\ell_{2}}^{\rm I}(t_{1})\rangle
-\langle\hat{c}_{j}^{\rm I,\dagger}(0)\hat{c}_{\ell_{2}}^{\rm I}(t_{1})\rangle\langle\hat{c}_{\ell_{3}}^{\rm I,\dagger}(t_{1})\hat{c}_{i}^{\rm I}(t)\rangle\langle\hat{c}_{l_{1}}^{\rm I,\dagger}(t_{1})\hat{c}_{\ell_{4}}^{\rm I}(t_{1})\rangle\nonumber\\
&-\langle\hat{c}_{j}^{\rm I,\dagger}(0)\hat{c}_{\ell_{2}}^{\rm I}(t_{1})\rangle\langle\hat{c}_{l_{1}}^{\rm I,\dagger}(t_{1})\hat{c}_{i}^{\rm I}(t)\rangle\langle\hat{c}_{\ell_{3}}^{\rm I,\dagger}(t_{1})\hat{c}_{\ell_{4}}^{\rm I}(t_{1})\rangle
-\langle\hat{c}_{j}^{\rm I,\dagger}(0)\hat{c}_{\ell_{4}}^{\rm I}(t_{1})\rangle\langle\hat{c}_{l_{1}}^{\rm I,\dagger}(t_{1})\hat{c}_{i}^{\rm I}(t)\rangle\langle\hat{c}_{\ell_{2}}^{\rm I}(t_{1})\hat{c}_{\ell_{3}}^{\rm I,\dagger}(t_{1})\big]\Big\}.
\end{align}
With the expression of free Green's function $\langle\hat{c}_{i}^{\rm I}(t)\hat{c}_{j}^{\rm I,\dagger}(0)\rangle=\delta_{ij}(1-n_{i})e^{-iE_{i}t}$ and $\langle\hat{c}_{j}^{\rm I,\dagger}(0)\hat{c}_{i}^{\rm I}(t)\rangle=\delta_{ij}n_{i}e^{-iE_{i}t}$,  we obtain the retarded Green's function
\begin{align}
\mathcal{G}_{ij}\left(t,0\right) &
=\Theta\left(t\right)\left[\mathcal{G}_{ij}^{>}\left(t,0\right)-\mathcal{G}_{ij}^{<}\left(t,0\right)\right]\nonumber\\
&=-ie^{-iE_{i}t}\delta_{ij}+\sum_{\ell_{1}}\sum_{\ell_{2}}\sum_{\ell_{3}}\sum_{\ell_{4}}it\widetilde{M}_{\ell_{1},\ell_{2},\ell_{3},\ell_{4}}e^{-iE_{i}t}\nonumber\\
&\times\Big[-\delta_{j,\ell_{4}}\delta_{\ell_{3},i}\delta_{\ell_{1},\ell_{2}}n_{\ell_{1}}+\delta_{j,\ell_{2}}\delta_{i,\ell_{3}}\delta_{\ell_{1},\ell_{4}}n_{\ell_{1}}+\delta_{\ell_{1},i}\delta_{j,\ell_{2}}\delta_{\ell_{3},\ell_{4}}n_{\ell_{3}}+\delta_{\ell_{1},i}\delta_{j,\ell_{4}}\delta_{\ell_{2},\ell_{3}}\left(1-n_{\ell_{2}}\right)\Big].
\end{align}
By defining
\begin{align}
\mathcal{C}_{ij}=\sum_{\ell_{1}}\sum_{\ell_{2}}\sum_{\ell_{3}}\sum_{\ell_{4}}\widetilde{M}_{\ell_{1},\ell_{2},\ell_{3},\ell_{4}}\big[-\delta_{j,\ell_{4}}\delta_{\ell_{3},i}\delta_{\ell_{1},\ell_{2}}n_{\ell_{1}}+\delta_{j,\ell_{2}}\delta_{i,\ell_{3}}\delta_{\ell_{1},\ell_{4}}n_{\ell_{1}}+\delta_{\ell_{1},i}\delta_{j,\ell_{2}}\delta_{\ell_{3},\ell_{4}}n_{\ell_{3}}+\delta_{\ell_{1},i}\delta_{j,\ell_{4}}\delta_{\ell_{2},\ell_{3}}(1-n_{\ell_{2}})\big],
\end{align}
the retarded Green's function is rewritten as
\begin{align}
\hat{\mathcal{G}}(t,0)=-ie^{-iE_{0}t }[1-\hat{\mathcal{C}}t]\approx-ie^{-i(E_{0}-i\hat{\mathcal{C}}) t}.
\end{align}
Here $\hat{\mathcal{G}}(t, 0)$ and $\hat{\mathcal{C}}$ are both 2-by-2 matrices, whose matrix elements are ${\cal G}_{ij}$ and ${\cal C}_{ij}$. Note that the symbol `$\approx$' represents exponentiation which produces the behavior of longtime dynamics as predicted in master equation. The Fourier transform of $\hat{\mathcal{G}}(t, 0)$ gives
$$
\hat{\mathcal{G}}(\omega)=\frac{1}{\omega-E_{0}+i \hat{\mathcal{C}}}.
$$
In order to obtain the pole of $\hat{\mathcal{G}}(\omega),$ we firstly diagnolize $\hat{\mathcal{G}}(\omega)$ and obtain two eigenvalues
$$
\hat{\mathcal{G}}(\omega)=U_g^{-1}\left(\begin{array}{cc}
\mathcal{G}_{\lambda_{1}}(\omega) & 0 \\
0 & \mathcal{G}_{\lambda_{2}}(\omega)
\end{array}\right) U_g,
$$
where $U_g$ is a unitary rotation.
Then $A_{1}(\omega)=-\frac{1}{\pi} \operatorname{\rm Im}\left[\mathcal{G}_{\lambda_{1}}(\omega)\right]$ and $A_{2}(\omega)=-\frac{1}{\pi} \operatorname{\rm Im}\left[\mathcal{G}_{\lambda_{2}}(\omega)\right]$ give two branches of spectral function as shown in Fig. 1 of the maintext.

\end{document}